\documentclass[fleqn,10pt,twocolumn]{wlscirep}
\usepackage[version=3]{mhchem}

\begin{document}

\title{Characteristics of ferroelectric-ferroelastic domains in N{\'e}el-type skyrmion host GaV$_4$S$_8$}

\author[1,*]{{\'A}d{\'a}m Butykai}
\author[1]{S{\'a}ndor Bord{\'a}cs}
\author[1]{Istv{\'a}n K{\'e}zsm{\'a}rki}
\author[2,3]{Vladimir Tsurkan}
\author[2]{Alois Loidl}
\author[4]{Jonathan D\"{o}ring}
\author[4]{Peter Milde}
\author[4]{Susanne C. Kehr}
\author[4,5]{Lukas M. Eng}

\affil[1]{Department of Physics, Budapest University of Technology and Economics and MTA-BME Lend\"{u}let Magneto-optical Spectroscopy Research Group, 1111 Budapest, Hungary}
\affil[2]{Experimental Physics V, Center for Electronic Correlations and Magnetism, University of Augsburg, 86135 Augsburg, Germany}
\affil[3]{Institute of Applied Physics, Academy of Sciences of Moldova, MD 2028, Chisinau, Republica Moldova}
\affil[4]{Institut f\"{u}r Angewandte Physik, Technische Universit\"{a}t Dresden, D-01062 Dresden, Germany}
\affil[5]{Center for Advancing Electronics Dresden cfaed, Technische Universit\"{a}t Dresden, 01062 Dresden, Germany}

\affil[*]{butykai@dept.phy.bme.hu}


\begin{abstract}
GaV$_4$S$_8$ is a multiferroic semiconductor hosting N{\'e}el-type magnetic skyrmions dressed with electric polarization. At T$_s$ = 42\:K, the compound undergoes a structural phase transition of weakly first-order, from a non-centrosymmetric cubic phase at high temperatures to a polar rhombohedral structure at low temperatures. Below T$_s$, ferroelectric domains are formed with the electric polarization pointing along any of the four $\left< 111 \right>$ axes. Although in this material the size and the shape of the ferroelectric-ferroelastic domains may act as important limiting factors in the formation of the N{\'e}el-type skyrmion lattice emerging below T$_C$=13\:K, the characteristics of polar domains in GaV$_4$S$_8$ have not been studied yet. Here, we report on the inspection of the local-scale ferroelectric domain distribution in rhombohedral GaV$_4$S$_8$ using low-temperature piezoresponse force microscopy. We observed mechanically and electrically compatible lamellar domain patterns, where the lamellae are aligned parallel to the (100)-type planes with a typical spacing between 100 nm-1.2 $\mu$m. We expect that the control of ferroelectric domain size in polar skyrmion hosts can be exploited for the spatial confinement and manupulation of N{\'e}el-type skyrmions.
\end{abstract}

\flushbottom
\maketitle
\thispagestyle{empty}

\section*{Introduction}
\label{sec1}

Ferroelectrics find numerous applications, amongst others, in non-linear electronic components, memory elements, electro-optical devices, piezoelectric transducers and actuators \cite{xu2013ferroelectric}. To minimize the electrical stray field energy, bulk ferroelectrics break up into domains, whose size distribution and relative orientation are of key importance for these applications. Furthermore, domain boundaries are interesting on their own due to their emergent functionalities \cite{catalan2012domain, meier2015functional}. For instance, domain walls can host itinerant electrons, attracting a lot of attention recently for domain wall conductivity \cite{seidel2009conduction,sluka2013free,schroder2014nanoscale,kagawa2009dynamics,meier2012anisotropic}. 

Besides proper ferroelectrics, in which the primary order parameter is the ferroelectric polarization, 'improper ferroelectrics' are also intensively studied. In these materials the electric polarization is often coupled to another degree of freedom, leading to new functionalities \cite{meier2012anisotropic}. Type-II multiferroics, where the polarization develops due to magnetic ordering, are well-known examples of improper ferroelectrics \cite{meier2015functional,cheong2007multiferroics,spaldin2005renaissance,kimura2003magnetic,kezsmarki2014one}. Other representative examples are the hexagonal manganites, in which polarization is driven by structural instability \cite{lilienblum2015ferroelectricity}. In these materials not only the bulk domains but also the domain walls are dressed with new properties such as interfacial magnetoelectricity \cite{kagawa2009dynamics} or anisotropic domain wall conductivity \cite{meier2012anisotropic}. 

Recent theoretical works have suggested that the Jahn-Teller effect, a symmetry-lowering structural distortion driven by the orbital degeneracy of localized charge carriers, may be responsible for a new type of improper ferroelectricity \cite{barone2015jahn, xu2015unusual}. In canonical Jahn-Teller materials with centrosymmetric crystal structure, such as KCuF$_3$, LaMnO$_3$, the distortion does not induce any electric polarization. In contrast, whenever the high-symmetry phase of a compound lacks inversion symmetry, a sizeable electric polarization may develop upon the Jahn-Teller transition. Theory suggests that, in strong contrast to structural transitions in proper ferroelectrics, the Jahn-Teller distortion is not suppressed in materials with a partially filled d shell, which in turn allows for simultaneous magnetic and ferroelectric orders, i.e. the emergence of multiferroic states \cite{barone2015jahn}. In our study, we report on the visualization and analysis of ferroelectric domain formation in the non-centrosymmetric Mott-insulator, GaV$_4$S$_8$, upon its ferroelectric phase transition driven by the cooperative Jahn-Teller effect \cite{ruff2015multiferroicity}. Using low-temperature piezoresponse force microscopy (PFM) and atomic force microscopy (AFM)  simultaneously ferroelectric and ferroelastic domains are visualized on different surfaces of GaV$_4$S$_8$.  

\begin{figure}
    \centering
    \includegraphics[width=\columnwidth]{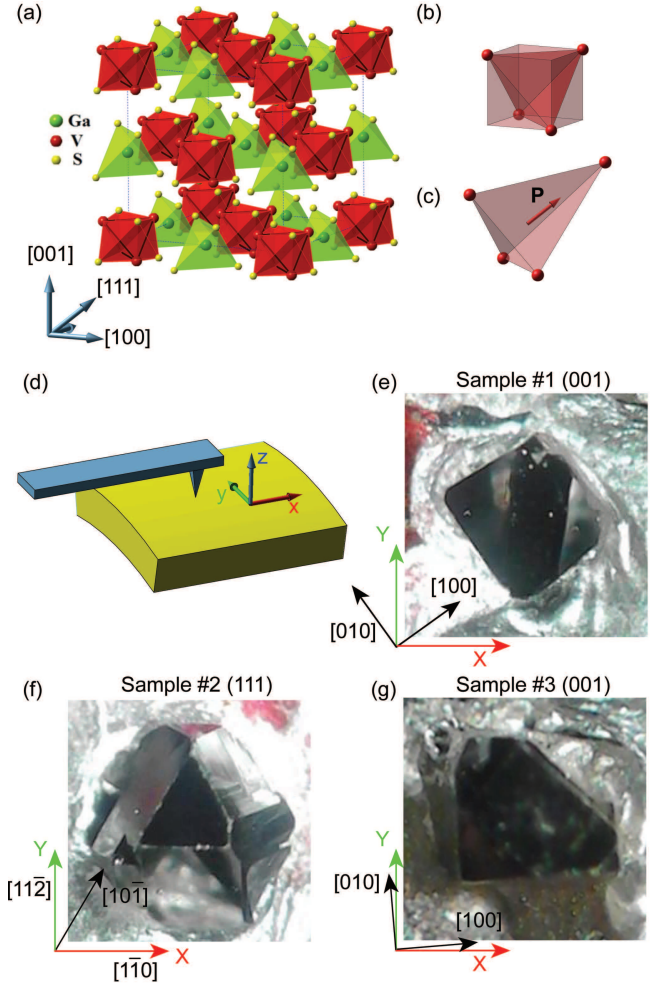}
    \caption{Panel (a) shows the crystal structure of GaV$_4$S$_8$. The red V$_4$S$_4$ clusters and the green VS$_4$ clusters are arranged in an fcc lattice. Panel (b) illustrates the tetrahedral arrangement of the vanadium sites within the V$_4$S$_4$ clusters in the cubic phase of the crystal. The rhombohedrally distorted V$_4$ cluster is shown in panel (c). The distortion is exaggerated for visibility. The local electric polarization developing upon the phase transition is indicated by a red arrow. Panel (d) demonstrates the schematic setup for out-of-plane piezoresponse force microscopy (PFM), with the lateral x and y scanning directions indicated by red and green arrows, respectively. Panels (e)-(g) present optical micrographs of the three different GaV$_4$S$_8$ single crystals used in this study, i.e. \#1 and \#3 with (001) surfaces, and \#2 exhibiting the (111) crystallographic surface. The typical dimensions of the cuboid-shaped crystals are approximately 1-2\:mm. The high-symmetry directions within the as-grown surfaces are indicated with reference to the x and y PFM scanning directions.}
    \label{fig:sample_setup}
\end{figure}

GaV$_4$S$_8$ belongs to the family of lacunar spinels, characterized by the vacancy of every second Ga-site in the unit cell as compared to the ordinary spinel structure. In an ideal spinel structure, the vanadium atoms would be arranged in a network of corner-sharing tetrahedrons, known as the phyrochore lattice. However, owing to the central cationic void in every second tetrahedron, the lattice decomposes into smaller and larger V$_4$ tetrahedral clusters, hence forming a breathing pyrochlore lattice \cite{okamoto2013breathing}. As shown in Fig. \ref{fig:sample_setup} (a), the smaller V$_4$ clusters constitute an fcc lattice. Each V$_4$ cluster carries a spin of S=1/2, since the triply degenerate $t_2$ cluster orbital is occupied by single unpaired electrons \cite{pocha2000electronic}. The orbital degeneracy is lifted via a cooperative Jahn-Teller distortion at $T_s = 42 \: \text{K}$ through the stretching of the lattice along any of the four $\left< 111 \right>$-type body diagonals in the pseudocubic system, as presented in Figs. \ref{fig:sample_setup} (b) and (c), reducing the crystal symmetry from non-centrosymmetric cubic (F$\bar{4}$3m) to polar rhombohedral (R3m) \cite{pocha2000electronic,hlinka2016lattice}. The electric polarization developing upon the Jahn-Teller transition is confirmed by recent pyrocurrent measurements \cite{ruff2015multiferroicity}. Dielectric spectroscopy \cite{wang2015polar} and specific heat measurements \cite{malik2013peculiar} imply that the ferroelectric transition in GaV$_4$S$_8$ is not displacive but of disorder-to-order type. Since the $\bar{4}$ symmetry is lost upon this transition and the T$_d$ point symmetry is reduced to its subgroup C$_{3v}$, the formation of four degenerate structural domains is allowed. The polarization in each structural domain is aligned parallel to one of the four $\left< 111 \right>$ axes of distortion. More specifically, local electric dipoles on all V$_4$ tetrahedra within a given domain point towards the same vertex, as shown in Fig.\ref{fig:sample_setup} (c). Since the inversion symmetry is already broken in the cubic F$\bar{4}$3m phase, +/- polarization domains cannot arise within a single inversion variant of the crystal. 

Note that throughout this study, the structural domains will be labelled by the direction of the local polarization, such as [111], [1$\bar{1}$$\bar{1}$], [$\bar{1}$1$\bar{1}$], [$\bar{1}$$\bar{1}$1], as displayed in Figs. \ref{fig:AFM_PFM_001} (g) and \ref{fig:AFM_PFM_111} (g). Although these domains represent only one of the two possible inversion variants, all arguments in this paper can be equally applied to the other inversion variant by reversing the sign of the polarization vectors.

In addition to its ferroelectric nature, GaV$_4$S$_8$ becomes magnetically ordered \cite{pocha2000electronic,muller2006magnetic,yadav2008thermodynamic} and thus multiferroic at T$_C$ = 13\,K \cite{kezsmarki2015neel,ruff2015multiferroicity}. Magnetization measurements revealed several magnetic phases \cite{nakamura2009low} that have been recently identified as the cycloidal state and the N{\'e}el-type skyrmion lattice (SkL) embedded in the ferromagnetic state \cite{kezsmarki2015neel}. The axial symmetry of the rhombohedral structure gives rise to an orientational confinement of the skyrmion cores in each domain along the magnetic easy-axis, which coincides with the direction of the Jahn-Teller distortion or equivalently the direction of the ferroelectric polarization. Due to the change in the orientation of skyrmion cores from domain to domain, the size of a consistent SkL must be limited by the size of the structural domain. Small-angle neutron scattering data indicate the lateral correlation length of the skyrmion lattice to be larger than 110\,nm \cite{kezsmarki2015neel}, though the actual value cannot be determined due to the resolution limit of the experiment. Clearly, the ultimate upper limit of the correlation length is set by the size of the structural domains. The local-scale investigation of these domains characterized by their size, shape and orientation is exactly the subject of the present study. Polar and magnetic orders are cross-coupled in GaV$_4$S$_8$, which leads to a different ferroelectric polarization in each magnetic phase \cite{ruff2015multiferroicity}. This suggests that N{\'e}el-type skyrmions in GaV$_4$S$_8$ carry electric polarization providing an intriguing potential for the detection and manipulation of individual skyrmions by means of local electric probes, such as scanning probe techniques or nano-circuits.

\section*{Results}

\begin{figure}
    \centering
    \includegraphics{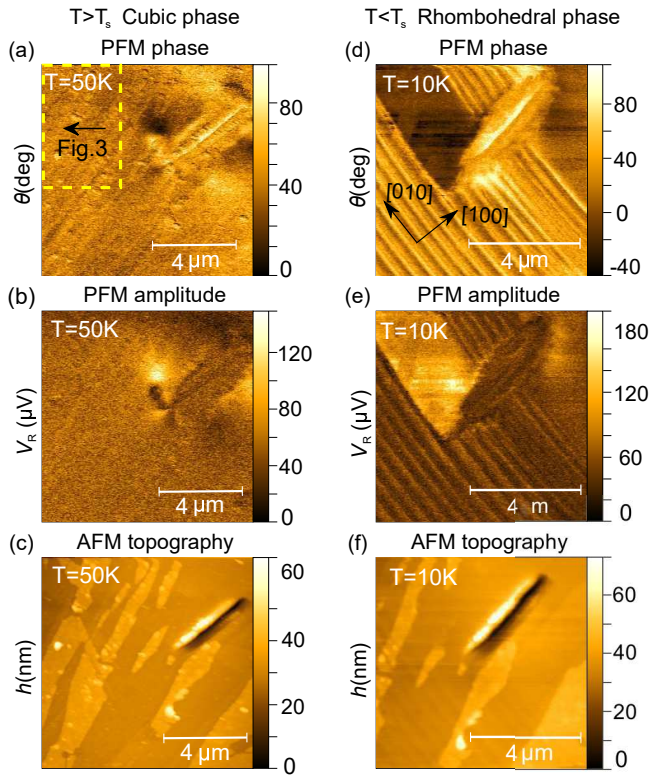}
    \caption{Simultaneous measurements of the PFM phase, PFM amplitude and AFM topography performed above [Figs. (a)-(c)] and below [Figs. (d)-(f)] the temperature of the structural phase transition. All images (a)-(f) stem from the same surface area on the (001) surface of Sample \#1, as clearly seen from the topographic AFM pictures (c) an (f). The crystallographic directions are indicated in panel (d) according to Fig. \ref{fig:sample_setup} (e). The yellow dashed rectangle in panel (a) marks the area of the measurements presented later in Fig \ref{fig:AFM_PFM_001}.}
    \label{fig:meas_001}
\end{figure}

Typical PFM micrographs recorded at the same position on the (001) plane of the first GaV$_4$S$_8$ sample, referred to as Sample \#1 in Fig. \ref{fig:sample_setup} (e), at $T= 50 \: \text{K}$ and $T= 10 \: \text{K}$ are shown in Fig. \ref{fig:meas_001}. The size of the scanned area was approximately 10\,$\mu$m $\times$ 10\,$\mu$m, the difference in the length scales at the two temperatures are due to the temperature-dependent response of the piezo actuator, which was calibrated at T=50\,K.

Since GaV$_4$S$_8$ has a non-centrosymmetric lattice even at high temperatures, in principle, antiphase or inversion domains could be present. In such antiphase domains, the magnitude of the converse piezoelectric effect is the same but has the opposite sign. However, this configuration is not appropriate to detect inversion domains, as the F$\bar{4}$3m crystal symmetry above the Jahn-Teller transition dictates the vanishing of the measured component of the converse piezoelectric tensor, $d^{[001]}_{zzz,c}$ (see supplementary information for details). Here, the direction indicated in the upper index refers to that of the AFM tip, the lower two entries denote the $zzz$-element of the tensor probed in the cubic ('c') phase of the crystal. This in accord with the structureless PFM images in Figs. \ref{fig:meas_001} (a)-(b) obtained at $50 \: K$. 

PFM images taken at $T=10\:K$, below the Jahn-Teller transition, reveal the appearance of stripe domains, that were analyzed both for the phase and amplitude dependence of the PFM signal [see Figs. \ref{fig:meas_001} (d) and (e), respectively]. As shown, the width of the ferroelectric domains ranges from 200 - 300\,nm. The same typical domain sizes have been found in other scanning areas of the (001) surface (see Supplementary Information for more images). Irregularities in the alternating domain structure are observed close to the center of the image due to a topographic defect. Static voltages up to $\pm$10\,V applied between the sample and the tip did not indicate any local polarization switching. Even though the ordered magnetic states were shown to carry sizable magnetoelectric polarization \cite{ruff2015multiferroicity}, the lateral resolution of the PFM measurement, approximately 50 nm, is not sufficient to resolve the polarization pattern accompanying the nanometric magnetic modulations.

\begin{figure}
    \centering
    \includegraphics{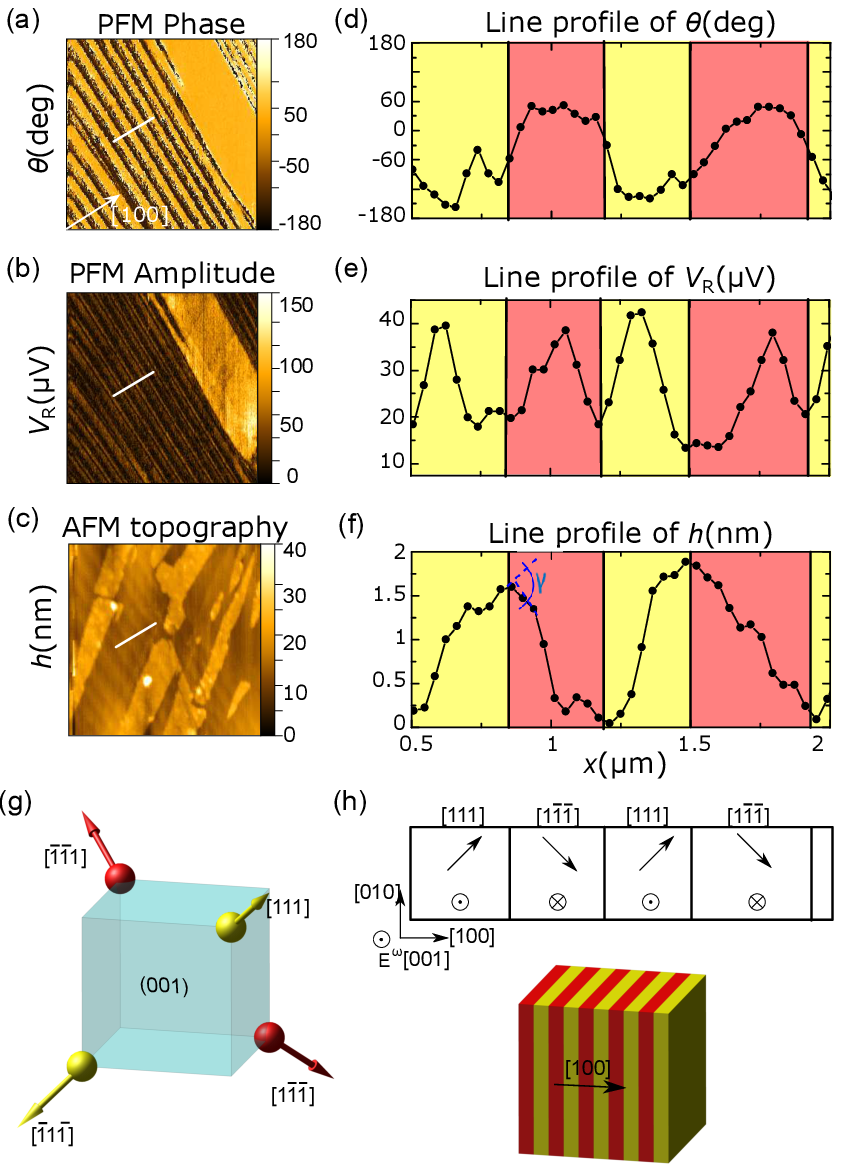}
    \caption{PFM phase (a) and amplitude (b) images along with the AFM topography image (c) taken on the (001) surface of Sample \#1 over the area marked by a yellow dashed rectangle in Fig. \ref{fig:meas_001} (a). Panels (d)-(f) present the corresponding profiles along the same line indicated by white bars in panels (a)-(c). The alternating yellow-red backgrounds of the line profiles indicate domains distinguishable by the AFM topography and the PFM measurements. The difference in the surface inclination angle in adjacent domains, denoted as $\gamma$, is displayed by a blue arc in panel (f). Panel (g) shows the polarization directions in the four rhombohedral domains in a single inversion variant of the crystal. Red and yellow colors represent the different sign of the out-of-plane PFM signal observable on the (001) surface. Panel (h) presents one of the two possible compatible solutions for the observed alternating structure along the line profile. The top image shows the in-plane and out-of plane polarization vectors with reference to the scanning directions and the alternating electric field applied by the PFM tip, denoted by $E^{\omega}$. The bottom image illustrates the three-dimensional structure of this solution with the color coding corresponding to panels (d)-(g).}
    \label{fig:AFM_PFM_001}
\end{figure}

The topography images of the scanned areas obtained from the dc component of the AFM signal were acquired together with the modulated PFM signals both at $T = 50 \: \text{K}$ and $T = 10 \: \text{K}$, as shown in Figs. \ref{fig:meas_001} (c) and (f), respectively. Besides the contrast observed in the PFM images, the topography image also reveals the presence of structural domains in the rhombohedral phase due to the different surface inclinations of the lattice in adjacent domains. The faint contrast seen in Fig. \ref{fig:meas_001} (f) appears more pronounced in an area without any topographic defects as in Fig. \ref{fig:AFM_PFM_001} (a) and (d). Note that such a saw-tooth height profile typical to the lamellar domain patterns in GaV$_4$S$_8$, was already observed in AFM micrographs by K{\'e}zsm{\'a}rki $\textit{et al}$. \cite{kezsmarki2015neel}.

The Jahn-Teller distortion leads to the appearance of a piezo-response, manifested by the $d^{[001]}_{zzz,r}$ component, where the entry 'r' stands for the rhombohedral phase of the crystal. Whereas the magnitude of this element is equal in all rhombohedral domains, its sign is determined by the sign of $P_z$, the polarization component normal to the probed surface. Thus, the four domains can be divided into two pairs of opposite $P_z$ component: [111],$ [\bar{1}\bar{1}1]$ and $[\bar{1}1\bar{1}]$,$[1\bar{1}\bar{1}]$. In the PFM phase signal, $\Theta$, a 180$^{\circ}$ contrast inversion is expected between these pairs, whereas, the PFM amplitude, $V_R$ is expected to be equal for all domains. The electromechanical coupling between the tip and the bottom electrode, however, results in an appreciable background signal (see Methods) affecting both the relative phase and amplitude of the PFM signals \cite{jungk2006quantitative}.  

Figures \ref{fig:AFM_PFM_001} (a)-(c) display the PFM phase and amplitude images along with the AFM topography image recorded in the area indicated by the yellow dashed rectangle in Fig. \ref{fig:meas_001} (a). The PFM amplitude and phase maps are displayed background-corrected as described in the Methods section. Figures \ref{fig:AFM_PFM_001} (d)-(f) demonstrate the profiles of the three channels along the same identical line cut out perpendicular to the domain boundaries, as marked by white bars in each figure. The alternating domain structure along the profile can be easily traced both in the AFM topography and PFM channels. As seen, the ferroelectric as well as ferroelastic domain boundaries are reflected by the reversal of the inclination angle in the topographic profile and by the 180$^{\circ}$ phase shift of the PFM signal and the minimum in its magnitude.

The measured piezo-response falls within the range of 20-40\,$\mu$V, corresponing to displacements of 6-12\,pm as determined from the calibration of the $z$-displacement of the AFM tip. The driving voltage was 10\,V$_{pp}$, therefore the magnitude of the $d^{[001]}_{zzz,r}$-element is estimated to be in the range of 1-2\,pm/V. 

The total inclination angle between the surfaces inferred from the AFM topography profile, indicated by dashed blue arcs in Fig \ref{fig:AFM_PFM_001}. (f), yields a value of approximately $0.58^{\circ} \pm 0.08^{\circ}$. This value is in good agreement with the rhombohedral angle of $59.66^{\circ}$, as measured by X-ray diffraction at $20 \: \text{K}$ \cite{pocha2000electronic}, which corresponds to $\gamma^{[001]}=0.58^{\circ}$ (for details, see SI).

\begin{figure}
    \centering
    \includegraphics{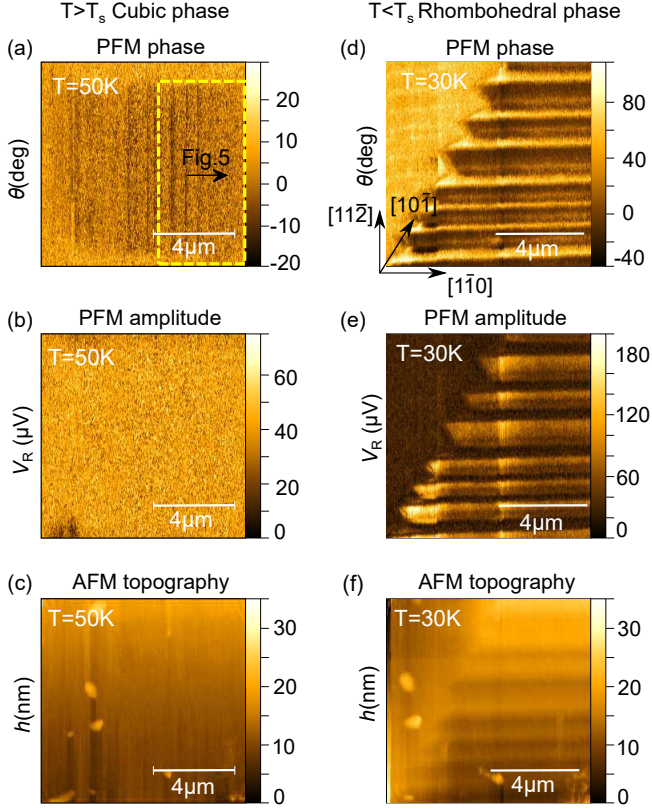}
    \caption{Simultaneous measurements of the PFM phase, amplitude and AFM topography performed above [Figs. (a)-(c)] and below [Figs. (d)-(f)] the temperature of the structural phase transition. All images (a)-(f) stem from the same surface area on the (111) surface of Sample \#2. The crystallographic directions are indicated in panel (d) according to Fig. \ref{fig:sample_setup} (f). The yellow dashed rectangle in panel (a) marks the area of measurements presented in Fig. \ref{fig:AFM_PFM_111}.}
    \label{fig:meas_111}
\end{figure}

In order to gain insight into the three-dimensional orientation of the domain walls, PFM measurements were carried out also on the (111) surface of GaV$_4$S$_8$ Sample \#2 [Fig. \ref{fig:sample_setup} (f)] at various temperatures. Figures \ref{fig:meas_111} (a)-(c) display the PFM phase, PFM amplitude and AFM images, respectively, measured at $T= 50 \: \text{K}$. Figures \ref{fig:meas_111} (d)-(f) present the same maps, captured over the same area with the sample cooled to $T= 30 \: \text{K}$. Similarly to the findings for the (001) plane, the PFM scans reveal stripe domain patterns in the rhombohedral phase, vanishing at temperatures above the structural phase transition. 

In contrast to the (001)-plane scans, the $d^{[111]}_{zzz,c}$ component of the converse piezoelectric tensor is allowed by the F$\bar{4}$3m symmetry, having opposite signs for the two antiphase domains. The vanishing PFM contrast in the high-temperature phase, however, suggests that the observed rhombohedral domains belong to the same inversion variant. Indeed, none of the high-temperature measurements on the (111) plane revealed any sign of antiphase domain boundaries (see Supporting Information). Therefore, it is reasonable to assume that the domain size of a pure inversion variant is large as compared to the scanned area, possibly extending over the whole crystal. The domain wall spacing measures approximately $1-1.5\mu m$, being much broader than those observed on the (001) plane. However, one additional stripe seems to be present within each larger domain, indicating a finer sub-structure. Also note that depending on the orientation of the domain walls the actual width of the domains might be smaller than the values determined from their (111)-sections observed here. The actual orientation of the domain walls will be discussed later.

\begin{figure}
    \centering
    \includegraphics{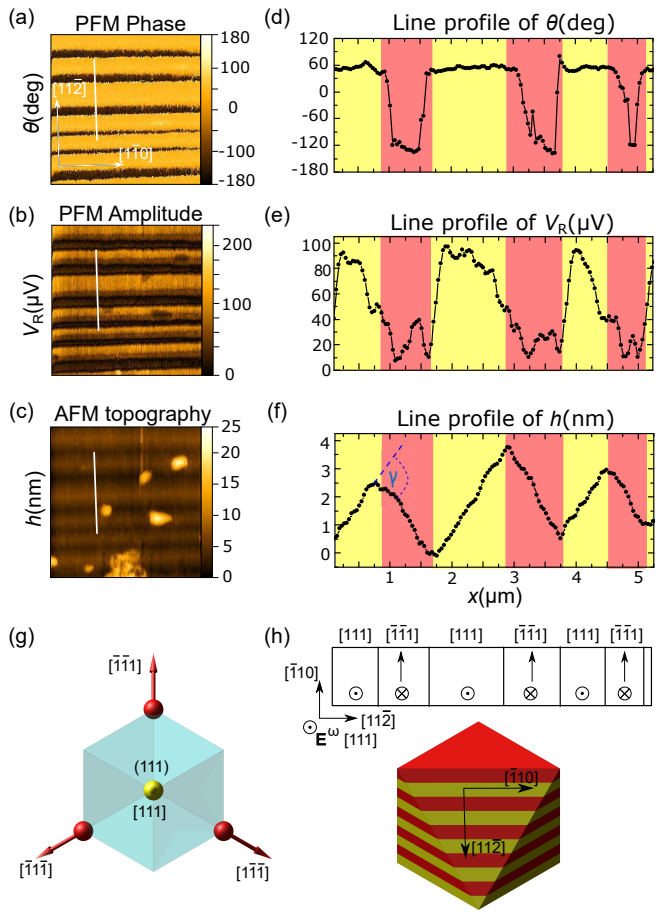}
    \caption{PFM phase (a) and amplitude (b) images along with the AFM topography image (c) taken on the (111) surface of Sample \#2 over the area marked by a yellow dashed rectangle in Fig. \ref{fig:meas_111} (a). Panels (d)-(f) present the corresponding profiles along the same line indicated by white bars in panels (a)-(c). The alternating yellow-red backgrounds of the line profiles indicate domains distinguishable by the AFM topography and PFM measurements. The difference in the surface inclination angle in adjacent domains, denoted as $\gamma$, is displayed by a blue arc in panel (f). Panel (g) shows the polarization directions in the four rhombohedral domains in a single inversion variant of the crystal. Red and yellow colors represent the different sign and magnitude of the out-of-plane PFM signal observable on the (111) surface. Panel (h) presents the only compatible solution that yields PFM contrast, as observed along the line profile. The top image shows the in-plane and out-of -plane polarization vectors with reference to the scanning directions. The bottom image displays the three-dimensional structure of this solution with the color coding corresponding to panels (d)-(g).}
    \label{fig:AFM_PFM_111}
\end{figure}

Figures \ref{fig:AFM_PFM_111} (a)-(c) present the background-corrected PFM phase and amplitude channels along with the AFM topography image recorded on the area marked by the yellow dashed rectangle in Fig. \ref{fig:meas_111} (a). Figures \ref{fig:AFM_PFM_111} (d)-(f) display the profiles along the white lines parallel to $[11\bar{2}]$, as indicated in the images in the left panels. The domain boundaries are clearly visible in all three channels. 

PFM measurements on the (111) plane in the rhombohedral phase are sensitive to the contrast in the converse piezoelectric constants between the unique $[111]$ domain and the three other domains $[1\bar{1}\bar{1}]$, $[\bar{1}1\bar{1}]$, $[\bar{1}\bar{1}1]$, whose $d^{[111]}_{zzz,r}$ elements are equivalent. In theory, the contrast between the piezo-response of the two distinguishable domain types yields a ratio of $3:-1$, where the negative sign indicates a $180^{\circ}$ shift in the PFM phase. 

The magnitude of the piezoelectric vibrations measured on the (111) plane is ~80-100\:$\mu$V for domain [111] and ~30-40\:$\mu$V for the other three domains. The corresponding value of $d^{[111]}_{zzz,r}$ is 2.5-5\:pm/V. This tensor element involves contributions from $d^{[001]}_{zzz}$, $d^{[001]}_{xxy}$ and $d^{[001]}_{xzx}$ (See Supplementary Information for details), which accounts for the enhancement in the PFM signal. 

The inclination of the (111) surface between the (111) domain and any of the three other domains is expected to be $\gamma^{[111]}=0.55^{\circ}$ from the degree of rhombohedral distortion determined by X-ray diffraction \cite{pocha2000electronic} (for details see SI). The opposing surface inclination in the alternating domains is clearly discernable from the AFM topography images [see Figs. \ref{fig:meas_111} (f) and \ref{fig:AFM_PFM_111} (a)-(d)], though the measured angle, $\gamma^{[001]}_{\textit{meas}}=0.38^{\circ} \pm 0.05^{\circ}$, is somewhat lower than the predicted value.

\section*{Discussion}

In a ferroelectric crystal the energy minimum is realized as an equilibrium between the cost of domain wall formation and the gain through the reduction in the stray field with increasing number of domains coexisting in the material \cite{shu2001domain, tsou2010compatible}. The mechanical continuity and electric neutrality of the interfaces between adjacent domains dictate a set of so-called compatibility equations \cite{shu2001domain}. In neighboring rhombohedral domains that belong to the same inversion variant, the polarization vectors parallel to the $\left<111\right>$-axes span approximately 109$^{\circ}$. For such rhombohedral domains, the compatibility criteria require that domain walls lay in (001)-type planes \cite{shu2001domain}.

A global energy minimum is typically achieved by the organization of the structural variants into alternating laminate patterns, as described by Tsou \textit{et al.} \cite{tsou2010compatible, tsou2011classification}. Regular domain patterns observed in most ferroelectrics can be understood and classified by their model. 

Indeed, all PFM images taken both on the (001) and the (111) planes of GaV$_4$S$_8$ reveal lamellar domain structures. In case of Sample \#1 (Figs. \ref{fig:meas_001} and \ref{fig:AFM_PFM_001}), the normal vectors of the observed domain walls correspond to a $\left< 10\xi \right>$-type pseudocubic direction of the crystal, where the $\xi$ component cannot be unambiguously determined from the measurement on this single surface. However,  $\xi=0$, corresponding to (100)-type domain walls is consistent with the compatibility criteria and separates structural domains originating from the same high-temperature inversion variant. In this case, the alternation of [111]- and [1$\bar{1}\bar{1}$]-type (Figs. \ref{fig:AFM_PFM_001} (g) and (h)) or [$\bar{1}1\bar{1}$]- and [$\bar{1}\bar{1}$1]-type variants (not shown) can be attributed to the observed lamellar domain structures. The two domains in any of the above pairs produce a PFM contrast measured on the (001) surface, represented by yellow and red colors in the figures. The measurement of the out-of-plane deformation does not allow for any further distinction between these two possible domain pairs. 
In the PFM measurements on Sample \#2, stripes parallel with the $\left< 1\bar{1}0 \right>$-type edges of the triangle-shaped (111) surface  have been evidenced [Figs. \ref{fig:meas_111} and \ref{fig:AFM_PFM_111}]. Such patterns can be explained again as compatible domain walls parallel to the (001) plane, which are projected to the (111) surface, as presented in Figs. \ref{fig:AFM_PFM_111} (g) and (h). The possible compatible domain pairs can be either [111] with [$\bar{1}\bar{1}1$] or [1$\bar{1}\bar{1}$] with [$\bar{1}1\bar{1}$]. However, PFM measurements on the (111) plane only reveal contrast in the former case, visualized by the yellow and red coloring in Figs. \ref{fig:AFM_PFM_111} (g) and (h). Within the dark area in the top left corner of the PFM amplitude image in Fig \ref{fig:meas_111} (e), an alternating structure of variants [1$\bar{1}\bar{1}$] and [$\bar{1}1\bar{1}$] must be present in order to maintain compatibility along the domain boundaries. Although these variants are indistinguishable via an out-of-plane PFM measurement, their presence can be traced in the PFM phase image [Fig. \ref{fig:meas_111} (d)] owing to the small contribution of the in-plane piezo-response to the measurement as a result of the non-uniform electric fields emerging near the tip.

We note that (110)-type domain walls could also account for similar patterns both in the (001) and (111) planes. However, the electric neutrality of such domain walls would require a 71$^{\circ}$ angle between the polarization of the adjacent domains \cite{shu2001domain}, which belong to high-temperature parent variants with an inverted structure. In the present experiments no sign of antiphase domains has been confirmed by PFM measurements above the temperature of the structural transition, moreover, the formation of lamellae of inversion variants at high-temperatures during the growth process seems rather unlikely. Thus our experiments show that the structural domain walls in GaV$_4$S$_8$ are parallel to (100)-type planes. Consequently, the actual distances between the (100)-type domain-walls in the bulk crystal are smaller by a factor of $\sqrt{3}$ than those observed on a (111) surface.

\begin{figure}
    \centering
    \includegraphics{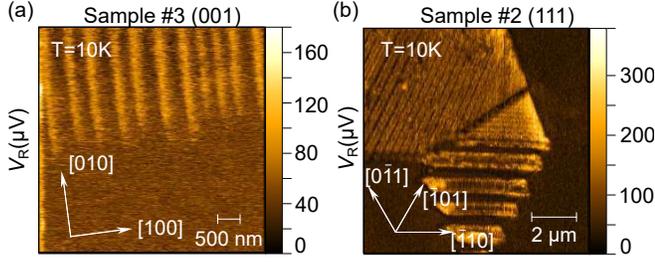}
    \caption{(a) Transition of alternating domain pairs into a uniform domain in Sample \#3 and (b) the junction of two different lamellar structures observed in Sample \#2. The matching of incompatible domains results in the deformation of the domain walls, producing needle-like endings, or irregular domain wall inclinations.}
    \label{fig:edges}
\end{figure}

Figures \ref{fig:edges} (a) and (b) present PFM amplitude images of the (001) and (111) surface of GaV$_4$S$_8$, respectively, captured in the rhombohedral phase close to the ends of the stripe domain patterns. Transitions of lamellar structures into another pair of alternating domains or to a uniform crystal variant may occur whenever the new phase starts developing from several nucleation centers such as structural defects. Epilayer plateaus may also serve as natural endings of lamellar structures. The resulting domain boundaries are prone to mechanical and electrostatic incompatibilities. Figure \ref{fig:edges} (a) reveals needle-like domain endings in an alternating structure as a consequence of mechanical stress and electrostatic repulsion induced at the incompatible domain walls \cite{salje1998needle}. Additionally, more complex domain patterns can arise as a manifestation of adjoining lamellae, as seen in Fig. \ref{fig:edges} (b). Thus, for future studies, it will be interesting to explore these domain boundaries more into depth, e.g. performing ~conducting AFM measurements in order to trace whether or not the compatibility criteria are broken, and whether these domain walls might possibly be rendered conductive. 

\section*{Conclusion}

We have detected ferroelectric-ferroelastic domains in the rhombohedral phase of GaV$_4$S$_8$ via simultaneous out-of-plane PFM and AFM topographic measurements. Lamellar domain structures have been observed and identified through alternating pairs of structural domains, belonging to a single inversion variant of the crystal. The magnitude of the piezoresponse measured on (001) and (111) faces of the crystals ranges within $\approx$1-5pm/V. Based on the mechanical and electric compatibility criteria we showed that the domain walls are parallel with (100)-type planes. The collection of PFM images supports that the compatibility criteria of non-charged domain walls are generally satisfied and also indicates the lack of antiphase domains in the crystals studied here. Notably, local incompatibilities have been encountered at the endings of the lamellar patterns, claiming for charged domain walls and polytype domain structures.  

The typical thickness of the observed lamella-like domains is in the range of 100 nm -1.2 $\mu$m (see SI for details). The wavelength of the cycloidal modulation and the diameter of N\'{e}el-skyrmions in GaV$_4$S$_8$ were reported to be $\approx$20\,nm \cite{kezsmarki2015neel}, at least one order of magnitude smaller than the thickness of the observed ferroelectric domains. Hence, the lateral correlation length of the skyrmion lattice may be limited by the structural domain size, however, the stability of the skyrmions is not affected by the thickness of the domains. On the other hand, the correlation length along the skyrmion tubes, i.e.~perpendicular to the 2-dimensional lattice, which were found to be of ~540\,nm \cite{kezsmarki2015neel}, is likely limited by the size of the rhombohedral domains. Since skyrmion cores are oriented along the polar axes of the ferroelectric domains, individual skyrmions are expected to be confined within the lamellar domains. We expect that engineering the ferroelectric domain patterns in polar skyrmion host materials provides a new route to tune, confine and manipulate skyrmions in bulk 3D materials.

\section*{Methods}
\label{sec2}
Single crystals of GaV$_4$S$_8$ were grown via chemical vapour transport method as reported elsewhere \cite{kezsmarki2015neel}. The typical dimensions of the cuboid-shaped crystals are 1-2\,mm [see Fig. \ref{fig:sample_setup} (e)-(g)]. The high quality of the samples has been confirmed by X-ray diffraction and specific heat measurements.

The mapping of the nanoscale topography and the local piezo-response were performed with a custom-made low-temperature AFM  \cite{doring2014near, doring2016low} on as-grown (001) and (111) surfaces of single crystalline GaV$_4$S$_8$ samples [see Figs. \ref{fig:sample_setup} (e)-(g)]. Commercial platinum-iridium coated silicon tips with a tip radius of about 30\,nm were used. An ac voltage of $\pm$5\,V at 22.3\,kHz was applied between tip and sample during contact-mode scans, with the cantilever deflection signal being demodulated at that same frequency by a lock-in amplifier to obtain the out-of-plane PFM amplitude and phase signals (note that the sample surface topography is recorded always simultaneously to PFM mapping). Due to the interferometric detection in our LT-PFM setup, only the out-of-plane $d_{zzz}$ element of the converse piezoelectric tensor was measured, where the $z$ direction always points normal to the sample surface in our notation [see Fig. \ref{fig:sample_setup} (d)].

In theory, structural domains with opposite signs of $P_z$, i.e.~the $z$-component of their polarization vector, feature a 180$^{\circ}$ phase contrast, whereas the amplitude of the PFM signal is proportional to the magnitude of $P_z$  \cite{jungk2009contrast,kalinin2007recent}. Besides the intrinsic PFM signal, the electromechanical coupling between the tip and the bottom electrode leads to an oscillation of the tip as well, manifested as a background signal of arbitrary phase with respect to the intrinsic PFM response. Hence, a PFM contrast between the domains emerges both in the phase and the amplitude channels, as seen in Figs. \ref{fig:meas_001} (d) and (e). The parasitic baseline signal is typically commensurate with the piezo-response of the sample \cite{jungk2006quantitative, xu2004effects, jungk2007consequences, bo2010influence}. 
The magnitude of the surface vibrations ($V_R$) originating exclusively from the converse piezoelectric effect can be estimated from the demodulated PFM amplitude and phase maps via background subtraction in the complex plane \cite{jungk2006quantitative}. 
In case of measurements performed on the (001) surface, the magnitude of the piezo-response is equal in all the four domains (for details, see SI), requiring that the complex PFM signal must be symmetric to zero. Hence, the complex components of the baseline, $V_{X0}$ and $V_{Y0}$, are determined as the spatial average values of the measured in-phase ($V_X$) and out-of-phase component ($V_Y$) of the PFM signal: $V_{X0}=\bar{V}_X$ and $V_{Y0}=\bar{V}_Y$. As a result, the background-corrected signal becomes symmetric to the origin of the complex plane. The magnitude and the phase of the surface piezoelectric vibration thus can be expressed as  \cite{jungk2006quantitative}: 

\begin{equation}
\begin{aligned}
    V_R=\sqrt{(V_X-V_{X0})^2+(V_Y-V_{Y0})^2}, \\
    \Theta=Arg((V_X-V_{X0})+i(V_Y-V_{Y0})).
\end{aligned}
\label{eq:100}
\end{equation} 

In case of experiments performed on the (111) surfaces, the magnitude of the probed piezo-response component of the [111]-domain, is expected to be three times larger as compared to the other three domains (see Supplementary Information). Therefore, the baseline correction in Equation \ref{eq:100} is employed using the modified $V_{X0}=\frac{V_{X+}+3V_{X-}}{4}$ and $V_{Y0}=\frac{V_{Y+}+3V_{Y-}}{4}$ complex background components, where $V_{X,Y+}$ and $V_{X,Y-}$ are the average of the maximal and minimal amplitudes in the X and Y channels, respectively. 

The PFM phase and amplitude images presented in Figs. \ref{fig:AFM_PFM_001} and \ref{fig:AFM_PFM_111} correspond to raw measurement data background corrected according to Eq. \ref{eq:100}, using the averaging described for the (001) and the (111) plane, respectively.

\bibliography{FE_ref}

\section*{Acknowledgement}
The authors thank J. Hlinka for the enlightening discussions, and the Helmholtz-Zentrum Dresden-Rossendorf, Germany for providing lab space, technical support, and the required cryogenics. The work was supported by the Deutsche Forschungsgemeinschaft
via the collaborative research center TRR 80 (Augsburg, Munich, Stuttgart), the Cluster of Excellence 'Center for Advancing Electronics Dresden (cfaed)', the Collaborative Research Center 'Correlated Magnetism: From Frustration to Topology' (SFB 1143) via TP C05, and through Project No. KE2068/2-1, by the Hungarian Research Funds OTKA K108918, PD 111756, and Bolyai 00565/14/11 as well as by the BMBF under Grants No. 05K10ODB and 05K16ODA.

\section*{Author Contributions}
J.D., A.B., S.B. performed the PFM measurements. A.B., S.B. analysed and interpreted the data. V.T. synthesized the crystals. A.B. wrote the manuscript with contributions from S.B. and I.K. S.B. supervised the project. The PFM setup was developed by S.K, J.D., L.M.E. All authors discussed and commented on the content of the paper.

\section*{Competing interests}
The authors declare no competing financial interests.

\end{document}